\documentclass[%
 reprint,
amsmath,
amssymb,
 aps,
 prd,
 floatfix,
]{revtex4-1}

\usepackage{latexsym}
\usepackage{longtable}
\usepackage{dcolumn}

\usepackage[pdftex]{graphicx}
\usepackage{bm}
\usepackage{hyperref}
\usepackage{mathtools}

\include{Steering_def}
\newcommand{\bqa}{\begin{eqnarray}}
\newcommand{\eqa}{\end{eqnarray}}
\newcommand{\bqs}{\begin{eqnarray*}}
\newcommand{\eqs}{\end{eqnarray*}}
\newcommand{\nll}{\nonumber\\}
\newcommand{\ds }{\displaystyle}
\newcommand{\sss}[1]{\scriptscriptstyle{#1}}
\newcommand{\cpl}{c_+}
\newcommand{\cmi}{c_-}

\newcommand{\ip}[1]{u\left({#1}\right)}             
\newcommand{\iap}[1]{{\bar{v}}\left({#1}\right)}    
\newcommand{\op}[1]{{\bar{u}}\left({#1}\right)}     
\newcommand{\oap}[1]{v\left({#1}\right)}            
\newcommand{\cows}{c^2_{\sss{W}}}
\newcommand{\siws}{s^2_{\sss{W}}}
\newcommand{\mzs}{M^2_{\sss{Z}}}
\newcommand{\mzl }{M_{\sss{Z}}}
\def\gz{\Gamma_{\sss{Z}}}
\newcommand{\mw}{M_{\sss W}}
\newcommand{\mz}{M_{\sss Z}}
\newcommand{\ml }{m_l}
\newcommand{\sqs}{\sqrt{s}}

\newcommand{\SANC}{\texttt{SANC}}

\newcommand{\WHIZARD}{\texttt{WHIZARD}}
\newcommand{\CalcHEP}{\texttt{CalcHEP}}

\newcommand{\stw}{s_{\sss{W}}  }
\newcommand{\ctw}{c_{\sss{W}}  }

\DeclarePairedDelimiter\aSp{\langle}{|}
\DeclarePairedDelimiter\bSp{[}{|}
\DeclarePairedDelimiter\SpA{|}{\rangle}
\DeclarePairedDelimiter\SpB{|}{]}

\newcommand{\vertsp}[1][]{\,#1\vert\, \mathopen{}}

\DeclarePairedDelimiterX\SpAIA[2]{\langle}{\rangle}{{#1}\vertsp[\delimsize]{#2}}
\DeclarePairedDelimiterX\SpAIB[2]{\langle}{]}{{#1}\vertsp[\delimsize]{#2}}
\DeclarePairedDelimiterX\SpBIA[2]{[}{\rangle}{{#1}\vertsp[\delimsize]{#2}}
\DeclarePairedDelimiterX\SpBIB[2]{[}{]}{{#1}\vertsp[\delimsize]{#2}}

\DeclarePairedDelimiterX\SpAIIA[3]{\langle}{\rangle}{{#1}\vertsp[\delimsize]{#2}\vertsp[\delimsize]{#3}}
\DeclarePairedDelimiterX\SpAIIB[3]{\langle}{]}{      {#1}\vertsp[\delimsize]{#2}\vertsp[\delimsize]{#3}}
\DeclarePairedDelimiterX\SpBIIA[3]{[}{\rangle}{      {#1}\vertsp[\delimsize]{#2}\vertsp[\delimsize]{#3}}
\DeclarePairedDelimiterX\SpBIIB[3]{[}{]}{            {#1}\vertsp[\delimsize]{#2}\vertsp[\delimsize]{#3}}

\newcommand{\hel}[5]{{}_{#1}{}_{#2}{}_{#3}{}_{#4}{}_{#5}}
\newcommand{\labhel}[6]{{}^{#1}_{{#2}{#3}{#4}{#5}{#6}} }

\begin{document}

\title{
  One-loop electroweak radiative corrections to lepton pair production
  in polarized electron-positron collisions}

\author{S. Bondarenko}
\email{bondarenko@jinr.ru}
\affiliation{%
  Bogoliubov Laboratory of Theoretical Physics, JINR, Dubna, 141980 Russia}

\author{Ya. Dydyshka}
\altaffiliation[Also at ]{Institute for Nuclear Problems, Belarusian State University,
  Minsk, 220006  Belarus}
\author{L. Kalinovskaya}
\author{R. Sadykov}
\author{V. Yermolchyk}
\altaffiliation[Also at ]{Institute for Nuclear Problems, Belarusian State University,
  Minsk, 220006  Belarus}
\affiliation{%
 Dzhelepov Laboratory of Nuclear Problems, JINR, Dubna, 141980 Russia
}%

\date{\today}

\begin{abstract}
This paper presents the high-precision theoretical predictions for $e^+e^- \to l^-l^+$ scattering.
Calculations are performed using the {\tt SANC} system.
They take into account complete one-loop electroweak radiative
corrections as well as longitudinal polarization of initial beams.
Reaction observables are obtained using the helicity amplitude
method with taking into account initial and final state fermion masses.
Numerical results are given for the center-of-mass energy range $\sqrt{s}=250-1000$~GeV
with various degrees of polarization.
\end{abstract}

\maketitle

\section{Introduction}

Planted experiments
(with/without polarization of the initial beams)
in high energy physics for electron-positron annihilation 
have been proposed with the  capability  of  precise  measurement,
such  as  the  International  Linear  Collider(ILC)~\cite{homepagesILC,Irles:2019xny,
  Moortgat-Picka:2015yla,Baer:2013cma,Accomando:1997wt,Battaglia:2004mw},
the $e^+e^-$ Future  Circular  Collider  (FCC-ee)
\cite{homepagesFCCee,Abada:2019ono,Abada:2019lih,Blondel:2019ykp,Blondel:2018mad},
the  Compact  Linear  Collider  (CLIC) \cite{homepagesCLIC,CLIC:2016zwp,Charles:2018vfv},
and the  Circular  Electron  Positron  Collider  (CEPC) \cite{homepagesCEPC}.

The theoretical accuracy for the future   $e^+e^-$ colliders
should be better than 0.5$\%$ \cite{Blondel:2019qlh}.
A first calculation of the corrections to the $e^+e^- \to \mu^-\mu^+$ process
was done by Passarino and Veltman \cite{Passarino:1978jh}.
Most of the theoretical works on lepton pair production (LPP) have been concerned
with next-to-leading order (NLO) electroweak (EW) radiative corrections (RCs)
(see e.g. \cite{Bardin:1981sv},\cite{Bardin:1980fe},
\cite{Akhundov:1984mp}, 
\cite{Berends:1987bg},
\cite{Bardin:1989tq},
\cite{Hollik:1988ii})~before the LEP era.

They were the development of basic codes 
incorporated to the standard LEP tools
such as {\tt TOPAZ0} \cite {Montagna:1998kp},
{\tt ZFITTER} \cite{Bardin:1999yd}, 
and {\tt ALIBABA} \cite{Beenakker:1990mb},\cite{Beenakker:1990ma}.
A comprehensive review of the underlying theory and methods
which have been used to create
these codes can be found in the monograph~ \cite{Bardin:1999ak}.

Polarized electron/positron beams are important to achieve the relative uncertainty
of a few per mille for  measurements of the total cross section and left-right asymmetry.
In the LEP era results were presented 
for the theoretical support of the polarized $e^+e^-$ annihilation,
see~\cite{Bardin:1979qy},  
\cite{Hollik:1980vf},
\cite{Bohm:1982hr},
\cite{Bohm:1983rn},
\cite{Kukhto:1983pv},
\cite{Hollik:1988ii},\cite{Grunewald:1999wn}. 
However, the mentioned investigations have not created a tool at
the same level  accuracy for the polarized beams.

There are three main $e^+e^-$ processes intended to be used
for the high-precision luminometry propose at flavour factories and future colliders:
Bhabha, lepton pair production $e^+e^- \to l^- l^+$ (LPP)
and photon pair production  $e^+e^- \to \gamma\gamma$ (PPP).

In the series of papers (\cite{Bardin:2017mdd}, \cite{sanc-ppp}) and this paper we recall
the above-mentioned three processes taking into account the one-loop EW RC and
longitudinally polarized $e^+e^-$ initial beams.

At the moment the most modern and widely used generators {\tt BABAYAGA}
\cite{CarloniCalame:2017ioy,Balossini:2006sd,CarloniCalame:2003yt}
and {\tt KKMC} \cite{Jadach:1999vf,Jadach:2013aha} with one-loop RCs for these three processes,
however, do not support the polarization of the initial beams.

For the unpolarized case we have already presented a comprehensive
comparison of the $e^+e^- \to f \bar{f}$ process with the results
of {\tt ZFITTER} for all the light fermion production channels in~\cite{Andonov:2002xc}.
In this work we report a brief description of the calculation
of the electroweak radiative corrections for the lepton (muon) pair production
focusing on the high energy region, including contributions
of the longitudinal polarization of the initial states.

In the special case of LPP $e^+ e^- \to \tau^- \tau^+$ reaction 
the decays of the $\tau$ lepton can be used to determine their polarization,
which gives extra information on the $Z\tau\tau$ vertex.
The polarization effects in this channel will be given in another paper.

In the future $e^+e^-$ collider program the optimized accelerator parameters are:
the center-of-mass (c.m.) energy $250$ GeV and higher 
and longitudinal electron $\pm 80\%$ and positron $0, \pm 30\%$
degree of polarization. Moreover it proposes a balance between
the polarization and the c.m. energy sets for optimal physics diversity.
    
In this study the relevant contributions to the cross section are calculated analytically
using the helicity amplitudes approach, which allows one to evaluate the contribution of
any polarization, and then obtain numerical result.
For the first time, the helicity amplitudes were used not only for the
Born-like parts but also for the hard photon bremsstrahlung contribution taking into account
the initial and final masses of the radiated particles.
The effect of polarization of the initial beams is carefully
analyzed for certain states. The angular and energy dependence are also considered.

There are many papers devoted to study of the $e^+e^- \to \mu^+ \mu^-$ channel at the one-loop
level with polarized effects in the initial state, see e.g.
\cite{Hollik:1980vf},
\cite{Bohm:1982hr}~
and references therein.
It is highly non-trivial to perform a tuned comparison of the numerical results,
since the authors not always present a complete list of input parameters.

We have performed  the high-precision tests at the tree level using
the electron-positron branch of the {\tt MCSANC} integrator~\cite{Bondarenko:2013nu},
\cite{Arbuzov:2015yja}
and the generator {\tt ReneSANCe}\cite{Sadykov:2020any} with the results of alternative codes.
The polarized Born and hard photon bremsstrahlung contributions were compared with
the corresponding values obtained with the help of the
{\tt CalcHEP}~\cite{Belyaev:2012qa} and {\tt WHIZARD} packages~\cite{Ohl:2006ae,
  Kilian:2007gr,Kilian:2014nya,Kilian:2018onl}.
The sum of virtual and soft photon bremsstrahlung contributions 
in the unpolarized case are compared with the {\tt AItalc-1.4}
code~\cite{Fleischer:2006ht}.

The numerical estimations are presented for the total
and differential cross sections in the scattering angle~$\cos\vartheta_l$,
and the relative corrections.
Also the left-right asymmetry $A_{LR}$ is given.

The paper is organized as follows.
Section~\ref{sec1} is
devoted to the expressions for the covariant (CAs) and helicity amplitudes (HAs)
for the Born, virtual and hard photon bremsstrahlung contributions.
The approach for the estimation of the polarization effects is also given.
Section~\ref{sec2} contains numerical results for the total and differential cross
sections as well as for relative corrections and $A_{LR}$ asymmetry.
The comparison with other computer codes are also given.
Finally, in Sec.~\ref{sec3} we present conclusions and outlook for the further work on LPP
process within the {\tt SANC} framework.

\section{Differential cross section}
\label{sec1}

The cross section of the generic process $e^+e^- \to ...$
of the longitudinally polarized $e^+$ and $e^-$ beams with the polarization degrees
$P_{e^+}$ and $P_{e^-}$, 
can be expressed as follows:
\begin{equation}
\sigma{(P_{e^+},P_{e^-})} = \frac{1}{4}\sum_{\chi_1,\chi_2}(1+\chi_1P_{e^+})(1+\chi_2P_{e^-})\sigma_{\chi_1\chi_2},
\label{eq1}
\end{equation}
where $\chi_{1(2)} = -1(+1)$ correspond to the particle $i$ with the left (right) helicity.

The complete one-loop cross section of the process can be split into four parts:
\bqa
\sigma^{\text{one-loop}} = \sigma^{\mathrm{Born}} + \sigma^{\mathrm{virt}}(\lambda)
+ \sigma^{\mathrm{soft}}(\lambda,\omega) + \sigma^{\mathrm{hard}}(\omega),
\nll
\eqa
where $\sigma^{\mathrm{Born}}$ is the Born cross section,
$\sigma^{\mathrm{virt}}$ is the contribution of virtual (loop) corrections,
$\sigma^{\mathrm{soft(hard)}}$ is the soft (hard) photon emission contribution
(the hard photon energy {$E_{\gamma} > \omega$}).
Auxiliary parameters $\lambda$ ("photon mass")
and $\omega$ are canceled after summation.

We apply the helicity approach to all the contributions.

The virtual (Born) cross section of the $e^+e^- \to l^-l^+$ process
\bqa
e^+(p_1, \chi_1) + e^-(p_2, \chi_2) \to l^-(p_3, \chi_3) + {l}^+(p_4, \chi_4)
\label{proc-eell}
\eqa
can be written as follows:
\bqa
\frac{d\sigma^{\mathrm{virt(Born)}}_{ \chi_1 \chi_2}}{d\cos{\vartheta_l}}
= \pi\alpha^2\frac{\beta_l}{8s}|\mathcal{H}^{\mathrm{virt(Born)}}_{\chi_1 \chi_2}|^2,
\eqa
where
\bqa
|\mathcal{H}^{\rm virt(Born)}_{ \chi_1 \chi_2}|^2 =
\sum_{\chi_3,\chi_4} |\mathcal{H}^{\rm virt(Born)}_{\chi_1 \chi_2\chi_3\chi_4}|^2,
\eqa
$m_l$ is the final lepton mass and $\beta_l=\ds\sqrt{1-\frac{4 m_l^2}{s}}$,
$\vartheta_l$ is the angle between the
final lepton $l^-$ and initial electron $e^-$.

The soft photon bremsstrahlung terms (initial state radiation - ISR, interference - IFI,
final-state radiation - FSR)
are factorized to the Born cross section as follows:
\bqs
&& \sigma^{{\rm soft},{\rm ISR}} =- \sigma^{\rm Born}\\
&&
\frac{\alpha}{\pi} Q^2_{e}
\Biggl\{
 \left(1+L_e \right)\ln\left(\frac{4\omega^2}{\lambda}\right)
     + L_e\left(1 + \frac{1}{2} L_e\right) + \frac{\pi^2}{3}
\Biggr\},
\\
&& \sigma^{{\rm soft},{\rm IFI}} = \sigma^{\rm Born}\frac{\alpha}{\pi} Q_eQ_{l}
\Biggl\{
 2\ln d_t \ln\left(\frac{4\omega^2}{\lambda}\right)
\\ \nonumber
&&       +\left[ 2\ln\left(1 + \frac{d_t^2}{st}\right)
       + \ln\left(-\frac{st}{d_t^2}\right) \right]\ln\left(-\frac{st}{d_t^2}\right) 
\\ \nonumber
&& + \left[2 J_l
   - \ln\left(1 - \frac{2m_l^2}{\beta_l^+ d_t}\right)\right]
         \ln\left(1 - \frac{2m_l^2}{\beta_l^+ d_t}\right)
\\ \nonumber
&&+ 2{\rm Li}_2\left(1 + \frac{2t}{\beta_l^+ d_t}\right)
- 2{\rm Li}_2\left(- \frac{d_t^2}{st}\right)
\\
&&   - 2{\rm Li}_2\left(\frac{\beta_l^-d_t}{(m_l^2 +t) - \beta_l d_t}\right)
\Biggr\}
-\Biggl\{ t \leftrightarrow u \Biggr\},
\\
&&\sigma^{{\rm soft},{\rm FSR}} = -\sigma^{\rm Born}
\\ &&
\frac{\alpha}{\pi} Q^2_{l} \frac{1}{\beta_l}
\Biggl\{
 \left[1+\left(1 - \frac{2m_l^2}{s} \right) J_l\right]
    \ln\Biggl(\frac{4\omega^2}{\lambda}\Biggr)
  + J_l
\\ &&
  +  \left(1 - \frac{2m_l^2}{s}\right)
  \left[\frac{1}{2} J_l^2+2\ln\left(-\frac{2\beta}{\beta_l^+}\right) J_l
    \right.
    \\
&&    \left.
  + 2 {\rm Li}_2\left(\frac{\beta_l^-}{\beta_l^+}\right)+\frac{\pi^2}{3}\right]
\Biggr\},
\eqs
where $L_e=\ln\left({m_{e}^2}/{s}\right)$, $\beta_l^\pm=1\pm\beta_l$,
$J_l = \ln(\beta_l^-)/\beta_l^+)$, 
 $d_I=m_l^2 - I, I=t,u$ .

The cross section for the hard photon bremsstrahlung 
\bqa
  e^+(p_1,\chi_1) + e^-(p_2,\chi_2) = l^-(p_3,\chi_3) + l^+(p_4,\chi_4) + \gamma(p_5,\chi_5).
\nll
\label{mainH-processes}
\eqa
is given by the expression
\bqa
\frac{d\sigma^{\mathrm{hard}}_{{ \chi_1}{ \chi_2}{}{}{} }}{ds'
        d\cos{\theta_4 }d\phi_4
        d\cos{\theta_5}}
=\alpha^3\frac{s-s'}{128\pi s^2}\frac{\beta_l'}{\beta_e}
|\mathcal{H}^{\mathrm{hard}}_{{ \chi_1}{ \chi_2}{}{}{}} |^2,
\eqa
where $s'=(p_3+p_4)^2$, $\beta_l'=\sqrt{1-4m_l^2/s'}$  and 
\bqa
|\mathcal{H}^{\mathrm{hard}}_{ \chi_1 \chi_2{}{}{}} |^2 =
\sum_{\chi_3,\chi_4,\chi_5} |\mathcal{H}^{\mathrm{hard}}_{{ \chi_1}{ \chi_2}{ \chi_3}{ \chi_4}{\chi_5}} |^2.
\eqa

Here $\theta_5$ is the angle between 3-momenta of the photon and positron,
$\theta_4$ is the angle between 3-momenta of the anti-muon $\mu^+$ and photon in the rest
frame of $(l^-l^+)$-compound, $\phi_4$ is the azimuthal angle of the $\mu^+$ in the rest frame of
$(l^-l^+)$-compound.

\subsection{Covariant amplitude for Born and virtual part}

The covariant one-loop amplitude (CA) corresponds to the result of the straightforward standard
calculation by means of {\tt SANC} programs and procedures of {\it all} diagrams contributing 
to the given process at the tree (Born) and one-loop levels.
It is represented in a certain basis, made of strings of Dirac matrices and/or external 
momenta (structures), contracted with polarization vectors of gauge bosons, if any. 
The amplitude also contains kinematic factors and
coupling constants
and is parametrized by a certain number of form factors (FF), which we denote by ${\cal F}$, 
in general with an index labeling the corresponding structure.
The number of FFs is equal to the number of structures. 

For the processes with non zero tree-level amplitudes the FFs have the form
\bqa
{\cal F} = 1 + k {\tilde{\cal F}}\,,
\eqa 
where ``1'' is due to the Born level and the term ${\tilde{\cal F}}$ with the factor
$k={g^2}/{16\pi^2}\,$
is due to the one-loop level.
After squaring the amplitude we neglect terms proportional to $k^2$.

Neglecting the masses of the initial particles
the covariant one-loop amplitude of the $e^+e^- \to l^-l^+$ process can be parametrized
by six FFs. If the initial-state masses were not ignored, we would have  ten structures
with ten scalar form factors and ten independent helicity amplitudes.
 

We work in the so-called $LQD$ basis, which naturally arises if the 
final-state fermion masses are not ignored.
Six form factors ${\cal F}_{\sss{LL,QL,LQ,QQ,LD,QD}}(s,t,u)$,
correspond to six Dirac structures. They are labeled
according to their structures.
A common expression for this CA in terms of ${\cal F}_{ij}$
was presented in~\cite{Andonov:2002xc}.
We recall it here to introduce the notations.
${\cal A}_{\gamma}$ is also described by a $QQ$ structure,
it is separated out for convenience
\bqa
&&{\cal A}_{\gamma}(s) =
i \,e^2\,\frac{\ds Q_{e} Q_{l}}{ \ds s}
Str_{\sss QQ}  {\cal F}_{\gamma}\,,
\label{ggNC}
\\
&&{\cal A}_{\sss{Z}}(s)=
i \,e^2\,\frac{\chi_{\sss{Z}}(s)}{s}
\nonumber\\ &&
\Big[ I^{(3)}_{e} \left(I^{(3)}_{l}~Str_{\sss LL}{\cal F}_{\sss{LL}} 
+                      \delta_{l}~Str_{\sss LQ}{\cal F}_{\sss{LQ}}\right)
\nonumber\\ &&
+ \delta_{e} \left ( I^{(3)}_{l}~Str_{\sss QL}{\cal F}_{\sss{QL}}
+ \delta_{l}~Str_{\sss QQ}{\cal F}_{\sss{QQ}}\right)
\nonumber\\ &&
+ I^{(3)}_{l}\left( I^{(3)}_{e} ~Str_{\sss LD}{\cal F}_{\sss{LD}}
                  +\delta_{e} ~Str_{\sss QD}{\cal F}_{\sss{QD}}\right)
\Big].
\nonumber
\eqa
We use the following notations for the structures
\bqs
Str_{\sss LL}&=& \gamma_\mu\left(1+\gamma_5\right)\otimes\gamma_\mu\left(1+\gamma_5\right),
\\
Str_{\sss QL}&=& \gamma_\mu \otimes \gamma_\mu \left( 1 + \gamma_5 \right),
\\
Str_{\sss LQ}&=& \gamma_\mu \left( 1 + \gamma_5 \right) \otimes\gamma_\mu, 
\\
Str_{\sss QQ}&=& \gamma_\mu\otimes\gamma_\mu,
\\
Str_{\sss LD}&=& \gamma_\mu{\left( 1 + \gamma_5 \right)}\otimes\left(- i m_l D_{\mu} \right),
\\
Str_{\sss QD}&=& \gamma_\mu \otimes \left( - i m_l D_{\mu} \right),
\eqs
where the symbol $\gamma_\mu\otimes\gamma_\mu$ denotes the short-hand notations
\bqa
\gamma_\mu\otimes\gamma_\nu=\iap{p_1}\gamma_{\mu}\ip{p_2}\op{p_3}\gamma_{\nu}\oap{p_4},
\eqa
and
\bqa
D_\mu=(p_4-p_3)_\mu.
\label{difference}
\eqa

Here and below   $\chi_{\sss{Z}}(s)$  is the $Z/\gamma$ propagator ratio:
\bqa
\chi_{\sss{Z}}(s)&=&\frac{1}{4\siws\cows}
\frac{\ds s}{\ds{s - \mzs + i\mzl\gz}}\,.
\label{propagators}
\eqa

We also use coupling constants
\bqs
Q_f\,,\quad I^{(3)}_f\,,\quad \sigma_f = v_f + a_f\,,\quad \delta_f = v_f - a_f\,,\\
\quad \stw=\frac{e}{g}\,,\quad 
\ctw=\frac{\mw}{\mz}.
\eqs

For more details see~\cite{Andonov:2002xc}.

\subsection{Helicity amplitude for virtual part}

As was stated we have six non-vanishing HAs.
They depend on kinematic variables, coupling constants and  six scalar form factors:
\bqs
    {\cal H}_{-++-} &=& - \cpl
    \left( Q_{e} Q_l {\cal F}_{\gamma}
    \right.
    \\
   &&\left.
    +{\chi_{\sss Z}(s)}\delta_{e}\left[
      \beta^{-}
      I^{(3)}_l{\cal F}_{\sss QL} +\delta_l {\cal F}_{\sss QQ}\right]\right),
    \\
      {\cal H}_{-+\pm\pm} &=&
      \frac{2\ml}{\sqs} \sin\vartheta_l
      \left(Q_{e} Q_l {\cal F}_{\gamma}
      \right.
      \\
  &&  \left.\hspace*{-2mm}
     +{\chi_{\sss Z}(s)}\delta_{e}\left[ I^{(3)}_l {\cal F}_{\sss QL}+\delta_l{\cal F}_{\sss QQ}
        +\frac{s}{2}
        \beta_l^2  I^{(3)}_l  {\cal F}_{\sss QD}\right] \right),
      \\
        {\cal H}_{+-\pm\pm} &=&
       - \frac{2\ml}{\sqs} \sin\vartheta_l
        \Bigl(Q_{e} Q_l {\cal F}_{\gamma}
        \\
  && +{\chi_{\sss Z}(s)}\Bigl[2I^{(3)}_{e} \left(I^{(3)}_l{\cal F}_{\sss LL}
          +\delta_l {\cal F}_{\sss LQ} \right)
          +\delta_{e} I^{(3)}_l {\cal F}_{\sss QL}
  \\ &&        
          +\delta_{e} \delta_l {\cal F}_{\sss QQ}
          +\frac{s}{2} \beta_f^2 I^{(3)}_l
          \Big( 2 I^{(3)}_{e} {\cal F}_{\sss LD}+\delta_{e} {\cal F}_{\sss QD}\Big) \Bigr] \Bigr),
        \\
 {\cal H}_{+--+} &=& - \cpl
          \Bigl(
          Q_{e} Q_l {\cal F}_{\gamma}
          \\
 && +{\chi_{\sss Z}(s)} \Bigl[ \beta^+ I^{(3)}_l
            \Big( 2 I^{(3)}_{e}   {\cal F}_{\sss LL}+\delta_{e} {\cal F}_{\sss QL}\Big)
            \\
 &&   +\delta_l  \left( 2 I^{(3)}_{e} {\cal F}_{\sss LQ}
            +\delta_{e}  {\cal F}_{\sss QQ} \right) \Bigr] \Bigr).
\eqs

The expression for the amplitude ${\cal H}_{-+-+}$ (${\cal H}_{+-+-}$) can be
obtained from the expression for ${\cal H}_{-++-}$ (${\cal H}_{+--+}$)
by replacing $\cpl \to \cmi, \beta^{-} \to \beta^{+}$.

Helicity indices denote the signs of the fermion spin projections to their momenta $p_1,p_2,p_3,p_4$, 
respectively. 

Where,
\bqs
\beta_l^\pm=1 \pm \beta_l,
~c_{\pm}=1 \pm \cos\vartheta_l,
\eqs
and the scattering angle $\vartheta_l$
is related to the Mandelstam invariants $t,u$:
\bqs
t  &=& m_l^2 - \frac{s}{2} (1 - \beta_l\cos\vartheta_l),
\\
u  &=& m_l^2 - \frac{s}{2} (1 + \beta_l\cos\vartheta_l ).
\eqs

\subsection{Helicity amplitudes for hard photon bremsstrahlung\label{HA_hard}}

We present the HAs for $e^+ e^- l^+ l^- \gamma \to 0$
($p_1+p_2+p_3+p_4+p_5 = 0$) process at any $s$, $t$ or $u$ channel, 
where $0$ stands for {\em vacuum}, and  all masses are not neglected.

We project all the  massive momenta with $p_i^2=m_i^2$ to the light-cone of photon $p_5$ and introduce associated ``momenta''
(auxiliary massless momenta)
\bqs
k_i = p_i - \frac{m_i^2}{2p_ip_5}p_5 = p_i - \frac{m_i^2}{2k_ik_5}k_5,
\\
k_i^2 = k_5^2 = 0,\   \text{ with }i=1,2,3,4.
\eqs
\bqs
k_5 = -\sum_{i=1}^{4}k_i = K p_5,\\
K = 1+ \sum_{i=1}^4\dfrac{m_i^2}{2p_i\cdot p_5}= 1+ \sum_{i=1}^4\dfrac{m_i^2}{2k_i\cdot p_5},
\\
p_5 = -\sum_{i=1}^{4}p_i = K' k_5,\\
K' = 1- \sum_{i=1}^4\dfrac{m_i^2}{2p_i\cdot k_5}  = 1- \sum_{i=1}^4\dfrac{m_i^2}{2k_i\cdot k_5} .
\eqs
The vector $k_5$ appears to be light-like, so we are left with ``momentum conservation''
of associated vectors.
The freedom in the light-cone projection choice corresponds to the arbitrariness of the spin
quantization direction.
We exploit it to make expressions compact.

 It is convenient to introduce the following notations
\bqs
{\cal D}^{ij}_I =\dfrac{Q_e Q_l}{I} +
\dfrac{g_e^i g_l^j}{I-M_Z^2+M_Z\Gamma_Z},
\eqs
where $I=s,s',~i=${\small \it L,R} and $j=${\small \it L,R}.

For massless particle with the light-like momentum $k_i$
we use the following notations and relations for spinors
\bqa
  &&\SpA{i} = u(k_i,+)=v(k_i,-),  \bSp{i} = \bar{u}(k_i,+)=\bar{v}(k_i,-), \nonumber \\ 
  &&  \SpB{i} = u(k_i,-)=v(k_i,+),  \aSp{i} = \bar{u}(k_i,-)=\bar{v}(k_i,+),
  \nonumber \\
&&\SpAIA{i}{j} = \SpAIA{k_i}{k_j}, 
\SpBIB{j}{i} = \SpBIB{k_j}{k_i},
\nonumber \\
&&\SpAIA{i}{j} = -\SpAIA{j}{i},
  \SpBIB{j}{i} = - \SpBIB{i}{j},
\nonumber \\
&& \SpAIA{i}{i} = 0, 
  \SpBIB{i}{i} =0,
    \SpBIB{j}{i}=\overline{\SpAIA{i}{j}}.\nonumber
\eqa


All non-vanishing amplitudes are obtained from  four amplitudes
by using CP and cross symmetries
\bqs
&&A^e\labhel{}{-}{-}{-}{+}{+} =
\\ &&
-\dfrac{m_{e}\SpAIA{4}{5} \SpBIB{1}{2}  }{K' \SpBIB{1}{5}\SpBIB{2}{5}} 
\bigg(  \dfrac{\SpBIB{2}{3}}{\SpAIA{1}{5}}{\cal D}_{s'}^{\sss LR}
        +\dfrac{\SpBIB{1}{3}}{\SpAIA{2}{5}}{\cal D}_{s'}^{\sss RR} \bigg),  
\\
&& A^e\labhel{}{-}{+}{-}{-}{+} =
\\ &&
-\dfrac{m_{l}\SpAIIB{5}{2}{1}}{K' \SpBIB{1}{5}\SpBIB{2}{5}}%
\bigg( \frac{\SpBIB{4}{1}}{\SpAIA{3}{5}}{\cal D}_{s'}^{\sss RL}
      + \frac{\SpBIB{3}{1}}{\SpAIA{4}{5}}{\cal D}_{s'}^{\sss RR} \bigg),
\\
&& A^e\labhel{}{-}{+}{-}{+}{+} =
\\ &&
-\frac{1}{\SpBIB{2}{5}} 
   \bigg[ 
     \frac{ m_{e}^{2}\SpBIB{1}{2}\SpBIIA{3}{5}{4}}{{\SpBIB{2}{5}} s_{51}}
           {\cal D}_{s'}^{\sss LR}
           + \frac{ m_{l}^{2} \SpAIIB{5}{2}{1}}{K'{\SpAIA{3}{5}}\SpBIB{4}{5}}
           {\cal D}_{s'}^{\sss RL}
   \\
&&   +  \frac{\SpBIB{1}{3}}{\SpBIB{1}{5}}
   \biggl( \frac{\SpAIIB{4}{2}{1} }{K'}+\SpAIIB{4}{5}{1} \biggr)
          {\cal D}_{s'}^{\sss RR}       
 \bigg],\nonumber
\\
&& A^e\labhel{}{+}{+}{-}{-}{-} =
\\ &&
- m_{e} m_{l} \SpAIA{1}{2} \Bigg(\frac{1}{s_{52}}
\bigg[
  \frac{  \SpBIB{4}{5} }{ \SpAIA{3}{5}}
       {\cal D}_{s'}^{\sss LL}
+ \frac{  \SpBIB{3}{5} }{ \SpAIA{4}{5}} {\cal D}_{s'}^{\sss LR}
   \bigg]
\\ &&
+\frac{1}{s_{51}}
   \bigg[
     \frac{  \SpBIB{4}{5} }{ \SpAIA{3}{5}} {\cal D}_{s'}^{\sss RL}
 +\frac{  \SpBIB{3}{5} }{ \SpAIA{4}{5}} {\cal D}_{s'}^{\sss RR} \bigg]
\Bigg),\nonumber
\eqs
where
\begin{eqnarray}
s_{ij} &= (p_i+p_j)^2.
\end{eqnarray}

Using the cross symmetry we can get
the lepton radiation amplitudes $A^l$
from the electron radiation radiation amplitudes $A^e$ in the following way
\bqs
&&    A^l \hel{ \chi_1}{ \chi_2}{ \chi_3}{ \chi_4}{\chi_5}(p_1, p_2, p_3, p_4,p_5)=
    \\
    &&    A^e \hel{ \chi_4}{ \chi_3}{ \chi_2}{ \chi_1}{\chi_5}(p_4, p_3, p_2, p_1,p_5)
~~\mbox{and}~~ m_e \leftrightarrow m_l.
\eqs

To obtain HA $\mathcal{H}$  with definite helicity, the spin-rotation
matrices $ C_{\xi_i}^{\phantom{a}\chi_i}$ should be applied for each index $\chi$
of external particles independently:
\bqa
\mathcal{H}_{... \xi_i ...} \hspace*{-1mm}=  2\sqrt{2}
C_{\xi_1}^{\phantom{a}\chi_1} \hspace*{-2mm}\dots C_{\xi_4}^{\phantom{a}\chi_4}
           \left(Q_e A^e_{... \chi_i ...} + Q_l A^l_{... \chi_i ...}\right).
           \nonumber
\eqa

\bqa
C_{\xi_i}^{\phantom{a}\chi_i}   &=& \left[\begin{matrix}\displaystyle
\frac{\SpBIB{i^{\flat}\!}{5}}{\SpBIB{i}{5}}
&\displaystyle
\frac{m_i\SpAIA{i^*}{5}}{\SpAIA{i^*}{i^{\flat}}\SpAIA{i}{5}}
\\\displaystyle
\frac{m_i\SpBIB{i^*}{5}}{\SpBIB{i^*}{i^{\flat}}\SpBIB{i}{5}}
&\displaystyle
\frac{\SpAIA{i^{\flat}}{5}}{\SpAIA{i}{5}}
\end{matrix}\right]
\\
&=&
\left[\begin{matrix}\displaystyle
\frac{\SpAIA{i^*}{i}}{\SpAIA{i^*}{i^{\flat}}}
&\displaystyle
\frac{m_i\SpAIA{i^*}{5}}{\SpAIA{i^*}{i^{\flat}}\SpAIA{i}{5}}
\\\displaystyle
\frac{m_i\SpBIB{i^*}{5}}{\SpBIB{i^*}{i^{\flat}}\SpBIB{i}{5}}
&\displaystyle
\frac{\SpBIB{i^*}{i}}{\SpBIB{i^*}{i^{\flat}}}
  \end{matrix}\right].
\nonumber
\eqa

\bqa
i &=& p_i = \{E_i,p_i^x,p_i^y,p_i^z\},\quad  p_i^2 =m_i^2,
\nll
i^*  &=& k_{i^*} = \{|\vec{p}_i|,-p_i^x,-p_i^y,-p_i^z\},\quad  k_{i^*}^2 =0,
\nll
i^{\flat} &=& k_{i^\flat} = p_i-\frac{m_i^2}{2p_i\cdot k_{i^*}}k_{i^*},\quad  k_{i^\flat}^2 =0.
\eqa

The CP symmetry allows one to obtain the flipped-helicity amplitudes
\bqs
    {\cal H}\hel{\chi_1}{\chi_2}{\chi_3}{\chi_4}{-} =
	-{\chi_1}{\chi_2}{\chi_3}{\chi_4}
	\overline{ {\cal H}} \hel{-\chi_1}{-\chi_2}{-\chi_3}{-\chi_4}{+} 
\eqs
with $L \leftrightarrow R$ in ${\cal D}$.

\section{Numerical Results and Comparisons}
\label{sec2}

In this section, we show numerical results for EW RC to
$e^+e^- \to {\mu}^-{\mu}^+$
scattering obtained by means of the {\tt SANC} system.
Comparison of our results for specific contributions at the tree level 
with {\tt CalcHEP} \cite{Belyaev:2012qa} and
{\tt WHIZARD}~\cite{Ohl:2006ae,Kilian:2007gr,Kilian:2014nya,Kilian:2018onl}
are  given.
The numerical results are completed with the estimation of the polarized effect
and evaluation of angular and energy distributions at the one-loop level.

We used the following  set of the input parameters
\begin{eqnarray}
\alpha^{-1}(0) &=& 137.03599976,
\\
M_W &=& 80.45150 \; \mathrm{GeV}, \quad M_Z = 91.1867 \; \mathrm{GeV},
\nonumber\\
\Gamma_Z &=& 2.49977 \; \mathrm{GeV}, \quad m_e = 0.51099907 \; \mathrm{MeV},
\nonumber\\
m_\mu &=& 0.105658389 \; \mathrm{GeV}, \quad m_\tau = 1.77705 \; \mathrm{GeV},
\nonumber\\
m_d &=& 0.083 \; \mathrm{GeV}, \quad m_s = 0.215 \; \mathrm{GeV},
\nonumber\\
m_b &=& 4.7 \; \mathrm{GeV}, \quad m_u = 0.062 \; \mathrm{GeV},
\nonumber\\
m_c &=& 1.5 \; \mathrm{GeV}, \quad m_t = 173.8 \; \mathrm{GeV}.
\nonumber
\end{eqnarray}
    
The $\alpha(0)$ and $G_\mu$ EW schemes are used in calculations.
All the results are obtained for  the c.m. energies $\sqrt{s}=250$, $500$ and $1000$~GeV
and for the following magnitudes of the electron $(P_{e^-})$ and the positron $(P_{e^+})$
beam polarizations:
\bqa
(P_{e^-}, P_{e^{+}})&& =
\label{SetPolarization} \\
&&(0,0),(-0.8,0),(-0.8,0.3),(0.8,0),(0.8,-0.3).
\nonumber
\eqa

\subsection{The comparison with another codes}

\subsubsection{The triple comparison of Born and hard photon bremsstrahlung cross sections}

First of all we compared the numerical results for
the polarized Born and hard photon bremsstrahlung cross section
with the ones obtained with the help of the {\CalcHEP} and {\WHIZARD}.
The agreement for the Born cross section was found to be exellent.

In the Table~\ref{Table:Hardh} the triple tuned comparison
between the {\SANC} (S) and the {\CalcHEP} (C) and {\WHIZARD} (W)
of the hard photon bremsstrahlung~(\ref{mainH-processes}) cross section
calculations are given.

\begin{table}[!h]
\centering	
\begin{tabular}{lcccc}
\hline
\hline
$P_{e^-},P_{e^+}$
& -1, -1       & 1, -1    & -1, 1 & 1, 1\\
\hline
\multicolumn{5}{c}{$\sigma_{e^+e^-}^{\text{hard}}$, fb, $\sqrt{s} = 250$, GeV}\\
\hline
S  & 169.0(1)  & 8802(1)   & 11263(1) & 169.0(1)  \\
C  & 169.8(1)  & 8824(2)   & 11294(2) & 169.8(1)  \\
W  & 167.3(1)  & 8802(1)   & 11261(2) & 168.4(1)  \\
\hline
\multicolumn{5}{c}{$\sigma_{e^+e^-}^{\text{hard}}$, fb, $\sqrt{s} = 500$, GeV}\\
\hline
S  & 47.38(1)  & 2314(1)   & 2899(1) & 47.38(1)  \\
C  & 47.63(3)  & 2318(1)   & 2905(1) & 47.55(4)  \\
W  & 46.81(1)  & 2313(1)   & 2900(1) & 46.94(1)  \\
\hline
\multicolumn{5}{c}{$\sigma_{e^+e^-}^{\text{hard}}$, fb, $\sqrt{s} = 1000$, GeV}\\
\hline
S  & 12.65(1)  & 624.7(1)   & 778.8(1) & 12.65(1)  \\
C  & 12.70(1)  & 626.1(1)   & 780.3(2) & 12.70(1)  \\
W  & 12.48(2)  & 624.7(1)   & 778.8(1) & 12.54(1)  \\
\hline\hline
\end{tabular}
\caption{The triple tuned comparison
between the {\SANC} (S) and the {\CalcHEP} (C) and {\WHIZARD} (W)
of the hard photon bremsstrahlung~(\ref{mainH-processes}) cross section
calculations.
}
\label{Table:Hardh}
\end{table}

The results are given within the $\alpha(0)$ scheme for c.m.
energies $\sqrt{s} = 250$, $500$ and $1000$~GeV, $\omega=10^{-4}$,
and the fixed 100\% polarized initial states in the total phase space.

The comparison demonstrates a very good (within 4-5 digits)
agreement with the above-mentioned codes.

\subsubsection {Comparison of virtual and soft photon\\ bremsstrahlung contributions}

We have obtained a very good agreement (six significant digits) in the comparison 
of the {\tt SANC} and {\tt AItalc-1.4}~\cite{Fleischer:2006ht} results for 
the unpolarized differential Born cross section and for the sum of 
the virtual and the soft photon bremsstrahlung contributions. 
The comparison was done for the different values of the
scattering angles $(\cos \vartheta$: from $-0.9$ up to $+0.9999)$.

\subsection{The Born, one-loop cross sections and relative corrections}

\subsubsection{The energy dependence}

In Tables~\ref{Table:delta_250}~-~\ref{Table:delta_1000}
the results of the Born cross sections,
weak contribution (weak) and complete one-loop contributions (EW)
as well as the relative corrections $\delta$ (\%)
for the c.m. energies $\sqrt{s}=250, 500, 1000$ GeV
and the set (\ref{SetPolarization}) of the polarization degree
of the initial particles in  the $\alpha(0)$ and $G_\mu$ EW schemes are presented.
The results were obtained without any angular cuts.

The relative corrections  $\delta$ (in \%) is defined as
\bqa
\delta = \frac{\sigma^{\text{one-loop}}(P_{e^-},P_{e^+})}{\sigma^{\text{Born}}(P_{e^-},P_{e^+})} -1.
\eqa

As it seen from the tables
the corrections for all considered c.m. energies, EW-schemes and
degrees of polarization are positive rather large and equal to
about 170-175\% for the c.m energy $\sqrt{s}=250$ GeV,
about 182-186\% for the c.m energy $\sqrt{s}=500$ GeV
and about 200-204\% for the c.m energy $\sqrt{s}=1000$ GeV in $\alpha(0)$
EW-scheme.
The calculations in the $G_{\mu}$ scheme reduce the RCs to the
about 5-6 \%.

The main impact to the one-loop corrections
is due to the QED contributions. It can be described by large
logarithms of the radiating particle masses ($\ln{s/m_l^2}$)
appeared for the collinear photons. The contribution of the collinear
photons is clearly seen from  Fig.~\ref{fig1}
where the rapid increasing of the cross section at small
angles of final muon ($|\cos\vartheta_{\mu}| \approx 1$) is observed.
The real-life experimental angular cuts can rapidly
reduce the QED radiative corrections and the whole cross section.

The degree of the initial particles polarization changes the magnitude
of the cross section, the minimal value achieved for unpolarized
beams and the maximum (from the set~\ref{SetPolarization}) 
for the $(P_{e^-}, P_{e^{+}})$ = (0.3,-0.8) ones. It can be useful to
increase the signal reaction.

\begin{table}[]
\caption{Born cross sections,
weak contribution (weak) and complete one-loop contributions (EW),
and relative corrections $\delta$ (\%)
for the c.m. energy $\sqrt{s}=250$ GeV
and the set (\protect\ref{SetPolarization}) of the polarization degree
of the initial particles in the $\alpha(0)$ and $G_\mu$ EW schemes.}
\label{Table:delta_250}
\centering
\begin{tabular}{lccccc}
\hline
\hline
$P_{e^+}$, $P_{e^-}$                   & 0, 0          & 0,-0.8       & 0.3,-0.8   &   0,0.8   & -0.3,0.8  \\
\hline \hline
$\sigma_{\alpha(0)}^{\text{Born}}$,   pb & 1.6537(1)     & 1.8040(1)    &  2.2572(1)  &  1.5034(1) & 1.8440(1) \\
$\sigma_{G_\mu}^{\text{Born}}$,   pb     & 1.7611(1)     & 1.9212(1)    &  2.4039(1)  &  1.6011(1) & 1.9638(1) \\
\hline
$\sigma_{\alpha(0)}^{\text{weak}}$, pb & 1.8360(1)       & 1.9447(1)    & 2.4261(1)   & 1.7273(1)  & 2.1271(1) \\
$\delta, \%$	                   & 11.03(1)        & 7.81(1)	    & 7.49(1)     & 14.89(1)  & 15.36(1) \\
\hline
$\sigma_{G_\mu}^{\text{weak}}$,   pb & 1.8547(1)   & 1.9614(1)          & 2.4466(1)   & 1.7480(1) & 2.1532(1)   \\
$\delta, \%$                     & 5.31(1)     &  2.10(1)	    & 1.78(1)     & 9.18(1)  & 9.64(1)  \\
\hline
$\sigma_{\alpha(0)}^{\text{EW}}$,   pb
                                & 4.534(1)    &	4.923(1)    & 6.115(1)      & 4.145(1)     & 5.047(1) \\
$\delta, \%$                    &174.2(1)     & 172.9(1)    & 170.9(1)      & 175.7(1)     & 173.7(1) \\
$\sigma_{G_\mu}^{\text{EW}}$, pb	& 4.728(1)    &	5.132(1)    &	6.376(1)     &  4.323(1)    & 5.263(1) \\
$\delta, \%$	                & 168.5(1)    &	167.1(1)    &	165.2(1)     &  170.0(1)    & 168.0(1) \\
\hline \hline
\end{tabular}
\end{table}

\begin{table*}[]
\caption{The same as in Tab.~\ref{Table:delta_250} for the c.m. energy $\sqrt{s}=500$ GeV.}
\label{Table:delta_500}
\centering	
\begin{tabular}{lccccc}
\hline
\hline
$P_{e^+}$, $P_{e^-}$                   & 0, 0         & 0,-0.8       &  0.3,-0.8   &   0,0.8   & -0.3,0.8  \\
\hline \hline
$\sigma_{\alpha(0)}^{\text{Born}}$,   pb & 0.40084(1)   & 0.43351(1)   &  0.54196(1) &  0.36820(1)  & 0.45215(1) \\
$\sigma_{G_\mu}^{\text{Born}}$,   pb     & 0.42689(1)   &	0.46167(1) &  0.57717(1) &  0.39211(1)  & 0.48152(1) \\
\hline
$\sigma_{\alpha(0)}^{\text{weak}}$,   pb &	0.44633(1) &	0.46766(1) &	0.58278(1) & 0.42501(1) & 0.52413(1) \\
$\delta, \%$	                     &	11.35(1)   &	7.88(1)	   &	7.53(1)    &   15.43(1)   & 15.92(1) \\
$\sigma_{G_\mu}^{\text{weak}}$,   pb     &	0.45095(1) &	0.47168(1) &	0.58768(1) & 0.43022(1) & 0.53067(1) \\
$\delta, \%$		             &  5.64(1)	   &	2.17(1)	   &	1.82(1)    & 9.72(1) & 10.21(1) \\
\hline
$\sigma_{\alpha(0)}^{\text{EW}}$,   pb & 1.145(1)     &	1.233(1)   & 1.531(1)      & 1.056(1) & 1.286(1) \\
  $\delta, \%$		            & 185.7(1)     &     184.5(1)   & 182.4(1)      & 186.8(1) & 184.5(1) \\
$\sigma_{G_\mu}^{\text{EW}}$,   pb     &	1.195(1)  &	1.287(1)   &	1.597(1)     & 1.102(1) & 1.342(1) \\
$\delta, \%$                       &	180.0(1)  &	178.8(1)   &	176.7(1)     & 181.1(1) & 178.8(1) \\
\hline
\hline
\end{tabular}
\end{table*}

\begin{table*}[]
\caption{The same as in Tab.~\ref{Table:delta_250} for the c.m. energy $\sqrt{s}=1000$ GeV.}
\label{Table:delta_1000}
\centering
\begin{tabular}{lccccc}
  \hline\hline
$P_{e^-}$, $P_{e^+}$                   & 0, 0          & 0,-0.8     & 0.3,-0.8   &   0,0.8   & 0.8,-0.3  \\
\hline
$\sigma_{\alpha(0)}^{\text{Born}}$,   pb &		0.099570(1) &	0.107474(1) &	0.134335(1) & 0.091666(1) & 0.112599(1) \\
$\sigma_{G_\mu}^{\text{Born}}$,   pb &		0.106038(1) &	0.114455(1) &	0.143061(1)  & 0.097620(1) & 0.119913(1) \\
\hline
$\sigma_{\alpha(0)}^{\text{weak}}$,   pb &       	0.11017(1)  &	0.11422(1)  &	0.14218(1)   & 0.10611(1) & 0.13103(1) \\
$\delta, \%$	                &               10.64(1)    &	6.28(1)	    &	5.85(1)      & 15.8(1) & 16.4(1) \\
$\sigma_{G_\mu}^{\text{weak}}$,   pb &      	0.11127(1)  &	0.11511(1)  &	0.14325(1)   & 0.10743(1) & 0.13269(1) \\
$\delta, \%$	                &                4.93(1)    &   0.57(1)	    &	0.13(1)      & 10.05(1) & 10.66(1) \\
\hline
$\sigma_{\alpha(0)}^{\text{EW}}$,   pb    & 0.3003(1) &	0.3223(1) & 0.3997(1)  & 0.2782(1) & 0.3392(1) \\
$\delta, \%$	                       &   201.6(1) &   199.9(1)   &	197.5(1) & 203.5(1) & 201.2(1) \\
$\sigma_{G_\mu}^{\text{EW}}$,   pb    &	        0.3137(1)  &	0.3367(1)  &	0.4176(1)   & 0.2907(1) & 0.3543(1) \\
$\delta, \%$	                  &              195.9(1)    &	194.2(1)    &	191.9(1)     & 197.8(1) & 195.5(1) \\
\hline\hline
\end{tabular}
\end{table*}

\subsubsection{The angular distributions}

\begin{figure}
  \begin{center}
    \includegraphics[width=85mm]{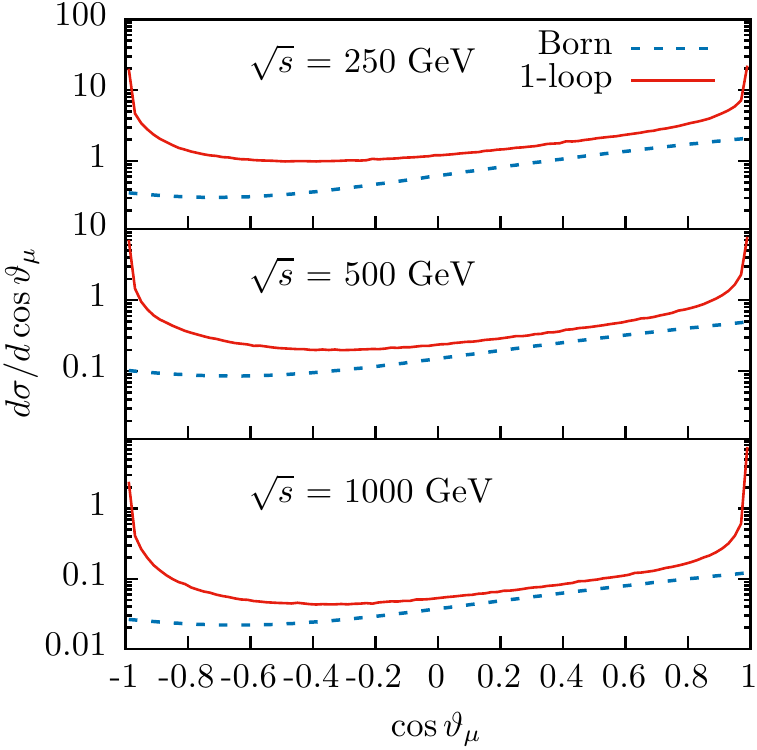}
    \caption{
      The Born (dashed line) and one-loop (solid line) differential cross sections
      of the $e^+e^- \to \mu^- \mu^+$ reaction for the c.m energies $\sqrt{s}=250$, 500, 1000 GeV.
      \label{fig1}
    }
  \end{center}
\end{figure}

In Fig.~\ref{fig1} the dependence of the muon angle of emission
is shown for the unpolarized Born and complete one-loop cross section
for c.m. energies $\sqrt{s} = 250$, $500$ and $1000$~GeV. As it seen the
Born distributions are rather smooth while the EW one-loop RCs
have large values at the small angles. As it was mentioned above, this is due to
the collinear emitted photons. Both distributions are asymmetric.

Since the polarization effects do not change the form
of the distributions, the only unpolarized cross section is shown.
The integrated values of the polarization effects are shown
in the Tables.~\ref{Table:delta_250}~-~\ref{Table:delta_1000}.

\subsection{The left-right asymmetry}
\label{La}

\begin{figure}
  \begin{center}
    \includegraphics[width=85mm]{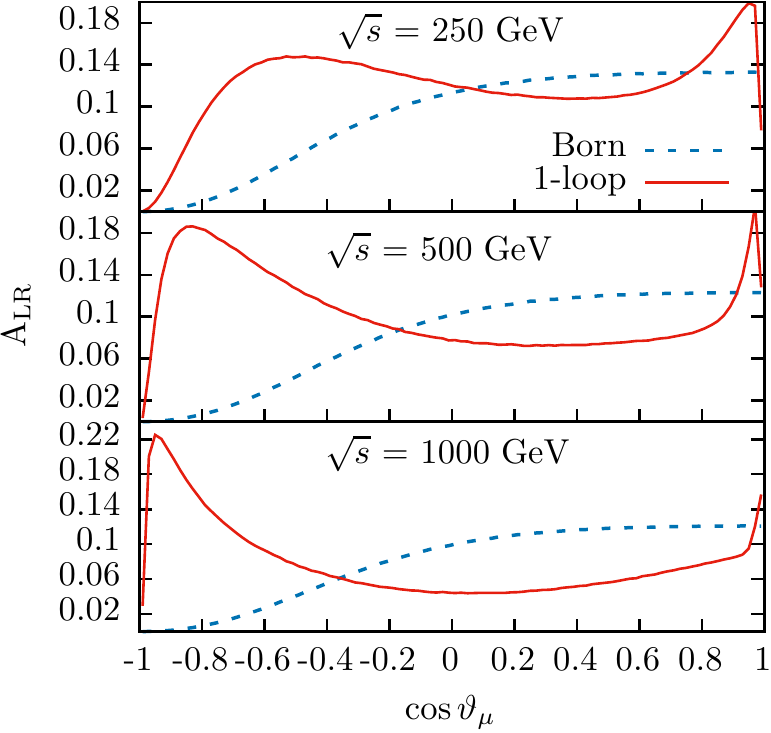}
    \caption{
      \label{ALR-cos4}
      The $A_{LR}$ asymmetries distribution dependence of the muon angle
      for Born (dashed line) and one-loop (solid line)
      contributions for three c.m. energies $\sqrt{s}=250, 500, 1000$ GeV.
      \label{fig2}
    }
  \end{center}
\end{figure}

In Fig.~\ref{fig2} the left-right asymmetry distributions are shown as a function of the
muon angle cosine. The $A_{LR}$ is defined in the following form
\bqa
A_{LR}=\frac{\sigma_{LR}-\sigma_{RL}}{\sigma_{LR}+\sigma_{RL}},
\eqa
where $\sigma_{LR}$  and $\sigma_{RL}$ are the cross sections for 100$\%$
polarized electron-positron $e^-_{L}e^+_{R}$ and  $e^-_{R}e^+_{L}$ initial states.

The $A_{LR}$ asymmetry distributions  for the Born and one-loop contribution are
shown for three c.m. energies $\sqrt{s}=250, 500, 1000$ GeV.

One sees that the EW RCs affect very strongly to the asymmetry.
The Born contribution to $A_{LR}$ has smooth dependance
on the $\cos\vartheta_{\mu}$
and equals to zero at $\cos\vartheta_{\mu}=-1$ and to the value $0.12-0.14$ depending
on the c.m energy.
The one-loop contribution has two maxima: first at $\cos\vartheta_{\mu}=1$ and
another one at
$\cos\vartheta_{\mu}=-0.6$ for $\sqrt{s}=250$ GeV,  $\cos\vartheta_{\mu}=-0.8$ for $\sqrt{s}=500$ GeV
and around the $\cos\vartheta_{\mu}=-1$ for $\sqrt{s}=1000$ GeV.

During the LEP era the $A_{LR}$ asymmetry (as well as the $A_{FB}$, $A_{FBLR}$ and final lepton polarization)
calculated at the $Z$ pole were used to measure experimentally the $\sin^2{\theta_{W}}$. The comprehensive
investigation of the one-loop contribution to the above-mentioned variables will be published
elsewhere.

\section{Conclusions and Outlook}
\label{sec3}

The theoretical description of  the $e^+e^- \to l^+l^-$ scattering 
taking into account the complete one-loop and high-order radiative corrections
is crucial for the luminosity monitoring at the modern and future $e^+e^-$ colliders.
Consideration of the beam polarization is a novel requirement
for the theoretical predictions for the $e^+e^-$ collisions at the energies of CLIC and ILC.

In the paper we have described the implementation 
of the complete one-loop EW calculations including the
hard photon bremsstrahlung contribution into the~{\SANC} framework.
It allows one to calculate the observables 
for the polarization processes of the lepton pair production.

In this study the relevant contributions to the cross section are calculated analytically
using the helicity amplitudes approach, which allows one to evaluate the contribution of
any polarization, and then estimated numerically.
For the first time the helicity amplitudes were  used not only for the
Born-like parts but also for the hard photon bremsstrahlung contribution  taking into account
the initial and final masses of the radiate particles.
The effect of polarization of the initial beams is carefully
analyzed for certain states. The angular and energy dependencies are also considered.

All contributions to the complete one-loop corrections, i.e. Born, virtual and
real soft- and hard photon bremsstrahlung were obtained using the helicity
amplitude approach. The independence of the form factors of the gauge parameters
was tested, the stability of the result from the variation of the soft-hard
separation parameter $\omega$ was checked.

The calculated polarized tree-level cross sections
for the Born and hard photon bremsstrahlung were
compared with the {\tt CalcHEP} and {\tt WHIZARD} results and
a very good (within 4-5 digits) agreement with the above-mentioned codes
was found.

Also we obtained a very good agreement (six significant digits) in the comparison 
of the {\tt SANC} and {\tt AItalc-1.4}~\cite{Fleischer:2006ht} results for 
the unpolarized differential Born cross section and for the sum of 
the virtual and the soft photon bremsstrahlung contributions. 

As a result, the polarization effects is significant and 
gives increase of the cross section at the definite initial degrees of polarization
compared to the unpolarized one.

We show that the complete ${\mathcal{O}}(\alpha)$ electroweak radiative corrections 
provide a considerable impact on the differential cross section and the
left-right asymmetry $A_{LR}$. Moreover, the corrections themselves are rather sensitive 
to polarization degrees of the initial beams and depend quite strongly on the energy.

Considering the $e^+e^- \to {l}^-{l}^+$ process as one
for luminometry propose,
one needs to take into account high-order effects, such as leading
multi-photon QED logarithms and mixed QCD-EW multi-loop corrections.
These corrections will be implemented in the future.

\section*{ACKNOWLEDGMENTS}
This work has been supported by the RFBR grant 20-02-00441.
We are grateful to Drs. A. Gladyshev and A. Sapronov
for the help in the preparation of the manuscript.

\providecommand{\href}[2]{#2}

\end{document}